\documentclass[twocolumn,showpacs,preprintnumbers,
superscriptaddress]{revtex4}
\usepackage{graphicx}

\begin{document}

\title{Probing Colour Separate States 
in $e^+ e^-$ annihilation at $Z^0$ pole}

\author{Shi-yuan Li}
\email{lishy@sdu.edu.cn}
\author{Feng-lan Shao}
\affiliation{
Department of Physics, Shandong University,
Jinan, Shandong 250100, People's Republic of China 
}
\author{Qu-bing Xie}
\email{xie@sdu.edu.cn}
\affiliation{
China Center of Advanced Science and Technology (CCAST),
Beijing 100080, People's Republic of China
}
\affiliation{
Department of Physics, Shandong University,
Jinan, Shandong 250100, People's Republic of China
}
\author{Qun Wang}
\affiliation{
Department of Physics, Shandong University,
Jinan, Shandong 250100, People's Republic of China 
}

\begin{abstract}

Hadronic events in $e^+ e^- \to Z^0$ 
are generated by a Monte-Carlo model including 
the production of colour separate states. 
These events are compared with those produced by 
the JETSET with default set of parameters  
where the colour connections are 
colour singlet chains. 
By selecting two-jet-like events and 
a sensitive observable, we find that these two 
kinds of colour connections lead to 
significant differences in the hadronic states.

\end{abstract}

\date{\today}


\maketitle


Hadronic processes in high energy collisions are generally described
by two distinct phases: the perturbative phase and the
non-perturbative hadronization one. 
The perturbative phase is well described
by perturbative QCD (PQCD) while the hadronization phase 
is non-perturbative and cannot be described
from first principle. Therefore a natural problem arises as to 
how to link these two phases, i.e. how to connect 
the partonic system produced by the PQCD evolution with 
the hadronization models, to give the final state hadrons.  
This problem is also beyond the capability of PQCD.

The parton cascade and hadronization models \cite{lund, webber} 
assume the string or cluster chain connections. 
They correspond to the partonic states where a 
color charge of one parton is specifically 
connected to its anti-color in
an accompanying parton. This approximation implies 
$N_C \to \infty$, thus with infinitely many colors the probability that
two (or more) partons have the same color is zero 
\cite{friberg}. The present hadronization models work well, 
which shows that the large $N_c$ limit reflects some 
feature of the real world.

However the colour structure of a multi-parton state 
is copious and complex at finite $N_C$
\cite{wangqunprd,wangq,wangqunold}.
This implies that the colour connections 
of a multi-parton system 
can be of many kinds. The 
ordinary one, which is used 
in the string and the cluster model \cite{lund, webber}, 
is just what we called the colour singlet chain (CC) state. 
The PQCD calculations show that, when projected     
onto the colour space of the final partonic state, 
such kind of colour states never appear with 
probability 1 at finite $N_C=3$. Its probability is 
shown to even decrease as the parton number of 
the partonic system increases \cite{wangqunprd,jinyi}. 
Another kind  of the colour connection is 
the colour separate (CS) state with 
several groups of gluons forming 
the colour singlet clusters each of which 
can hadronize independently \cite{friberg,wangq,tsj,lonnblad}. 
Such a possibility can be demonstrated 
by a simple example in 
$e^+e^- \rightarrow q \bar q + n g $
when two gluons have 'opposite' colours 
(e.g. $\bar r b$, $\bar b r$), 
they can form a colour singlet cluster by a closed string. 
When the parton number is large 
\footnote{as calculated by JETSET, the average numbers of
gluons are: $\sqrt{s}=91\; {\rm GeV}$, 
$\langle n \rangle\sim 6;~ \sqrt{s}=200\; {\rm GeV}, 
\langle n \rangle\sim 9; \sqrt{s}=1\; {\rm TeV}, 
\langle n \rangle \sim 17$.}, 
it is inevitable that 
more and more gluons have chances 
to form the CS states \cite{wangq,wangqunold}.
These two kinds of colour connections 
at the hard-soft interface 
correspond to different string/cluster configurations 
\cite{wangqunprd, wangq}
and may lead to differences in hadronic states 
through the subsequent hadronization process.  
In this sense the probability for any kind of colour connections 
depends on the non-perturbative QCD (NPQCD) mechanism. 
The cross sections of the CC and CS states calculated in the 
PQCD framework \cite{wangqunprd,wangq,wangqunold,jinyi} 
are not the final answer.

To study the effects of the CS states 
on the hadronic events, we need phenomenological models.  
The straightforward way is to modify the 
event generators, e.g. JETSET, by putting the  
CS states as one way of the colour connections. 
One can use the modified generators to give 
the hadronic states to see if there are 
any deviations from those produced by JETSET with the 
default colour connections (i.e. the CC connections). 
Recently we proposed a CS model \cite{wangq} 
based on JETSET 7.4 by replacing the 
default colour connections 
with the CS-allowed connections. 
Hereafter we also use CS and CC to refer to 
the modified JETSET and the default JETSET respectively. 
The simulated results show that 
there are no significant differences 
between the observables for the CS 
and CC unbiased events \cite{wangq,shaofenglan}.
This result is not hard to understand considering that  
for unbiased events the observables describing 
global properties are mainly
determined by PQCD and they are not much sensitive 
to the hadronization details.  
This property is nothing but the so-called  
local parton-hadron duality \cite{bo}. 
Therefore we have to look for possible sensitive 
observables defined for a specific set of events. 
We know that the baryon-antibaryon  
rapidity correlation is sensitive to the hadronization
models \cite{si}. It has been investigated 
if this correlation is also sensitive 
to the way of colour connections. 
But the answer is negative \cite{shaofenglan}. 
Considering that in a CS state 
only the phase space region around the 
glueball cluster and the two string pieces 
in its neighbourhood  
is much different from that of the corresponding CC state, 
the sensitive observables may exist in 
a specified window of the phase space. 

Based on the above analysis, in this paper we propose to 
select two-jet-like events in $e^+ e^- \to Z^0 \to {\rm hadrons}$ 
as an effective probe of the CS states.


We use the CS model in Ref. \cite{wangq}. 
In the model, we proposed a method to estimate the 
rate of the CS states for a group of partons produced 
in a hadronic event in high energy electron-positron 
collisions. We implement this model into a Monte-Carlo 
program based on JETSET 7.4. In the program we replace 
the default way of colour connections for the partonic 
system in JETSET 7.4, i.e. the CC connections, with the 
CS-allowed connections. 
The CS clusters formed by two or more gluons in a CS state
also hadronize in the usual way as closed strings \cite{lund}. 
We use the T-measure to weight 
the CS states \cite{wangq,shaofenglan}.

In this paper, a two-jet-like event at the partonic level 
is refered to that with an energetic back to back 
quark-antiquark pair and some associated soft gluons 
whose energies are smaller than a definite value $E_0$. 
In this case the thrust axis is 
approximately along the quark's momentum. 
Therefore in this paper 
we define the {\it longitudinal} and {\it transversal} directions 
with respect to the thrust axis. 
In the following numerical analysis, 
we require that the energies of all 
gluons including those appearing in the intermediate stage 
of the parton shower process are less than 2.5 GeV. 
This is to eliminate the possibility that the energetic gluons 
would significantly change the shape of a two-jet-like event.

We know that the larger the number of gluons in an event, 
the more possible does the CS state occur. 
To enlarge the effect, we only consider those events 
where the initial quark pair produced by 
the electroweak process is of $u$, $d$ or $s$ type 
because the number of gluons in heavy quark events 
is suppressed.

\begin{figure}
\includegraphics[scale=0.58]{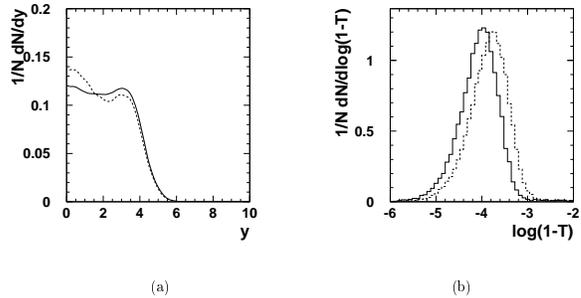}
\caption{The two-jet-like events selected at the partonic level.  
(a) Rapidity distribution of hadrons.   
(b) Thrust distribution. 
The solid line is for the CC events and 
dashed line for the CS ones, the same convention 
is implied for other figures.}
\label{a1} 
\end{figure}

\begin{figure}
\includegraphics[scale=0.58]{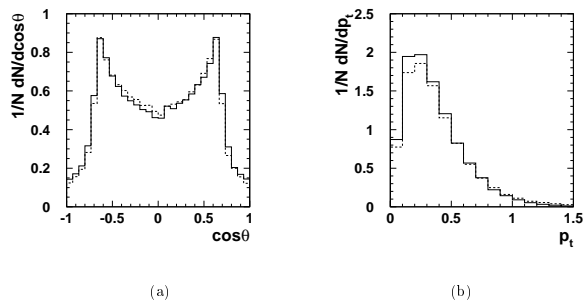}
\caption{ The two-jet-like events selected at the partonic level.
(a) The polar angle (the angle between
the momentum of a hadron and the
thrust) distribution of hadrons. (b)
The transversal momentum distribution of hadrons. }
\label{a2}
\end{figure}

For these events two different ways of colour connections 
really lead to differences in the hadronic states. 
Figs. \ref{a1} are the rapidity and thrust distributions 
of the hadronic states. One can see that 
the CS events have more particles with small rapidity 
and have smaller thrusts  than the CC ones.
The reason for this is that the formation of the closed strings 
in the CS events brings more momentum from the 
longitudinal to the transversal direction compared 
to the CC events. This point can be further strengthened 
by the polar angle and the transversal 
momentum distributions for the 
hadrons inside the rapidity window $|y|<0.8$, 
see Figs. \ref{a2}. The average transversal momentum 
of the hadrons in the CS events 
are larger than those in the CC events. 
There is a similar trend for the charged particle 
that there are more charged particles in the CS events than 
in the CC ones inside the central rapidity region.

\begin{figure}
\includegraphics[scale=0.58]{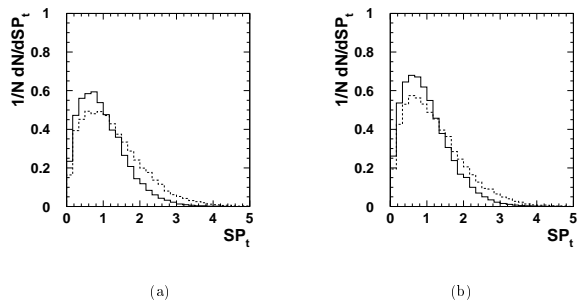}
\caption{ The distribution of $SP_t$ for the two-jet-like 
events selected at the partonic level; 
(a) $\sigma =$ default value; 
(b) $\sigma =0.29$.}
\label{a4}
\end{figure}

Furthermore we define the following observable:
\begin{equation}
SP_t=\sum_i|\vec{p}_{ti}|,\;\; \forall |y_i|<y_0 \; .
\end{equation}
where the $\vec{p}_{ti}$ and $y_i$ are 
the transversal momentum and rapidity of
a final state hadron in an event respectively. 
In the numerical calculation we take $y_0=0.8$. 
This observable is introduced to accumulate  the above 
differences of the CS and CC events. 
Fig. \ref{a4} (a) showes $SP_t$ is a sensitive observable to 
the way of the colour connections at the hard-soft interface.

\begin{figure}
\includegraphics[scale=0.55]{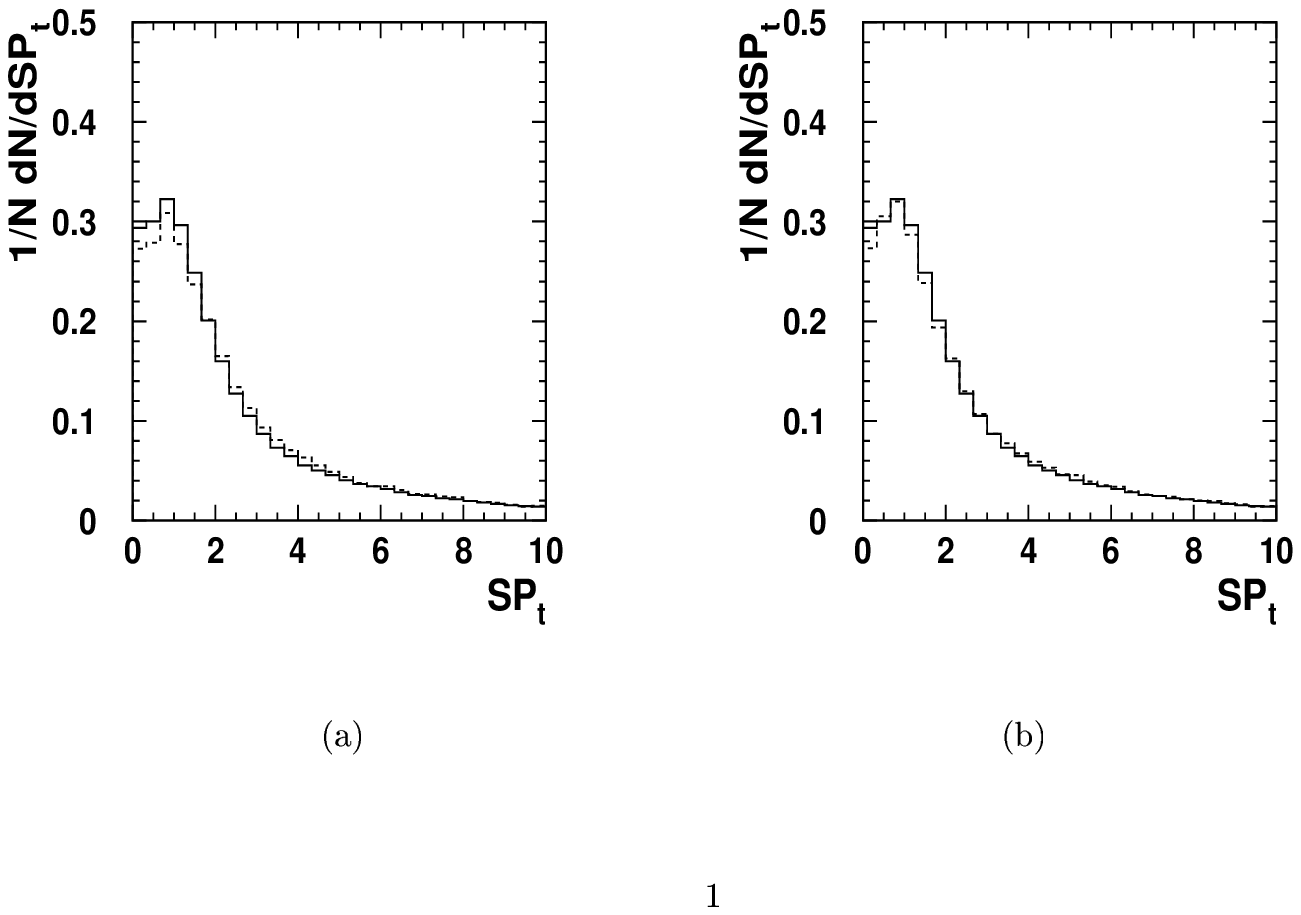}
\vskip -0.6cm
\caption{ $SP_t$ distribution for unbiased events; 
(a) $\sigma =$ default value; 
(b) $\sigma=0.29$. }
\label{a5}
\end{figure}

\begin{figure}
\includegraphics[scale=0.36]{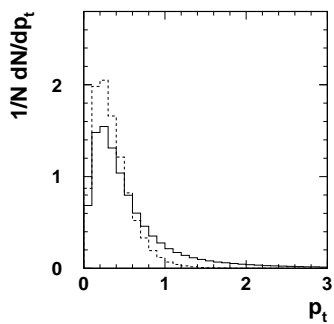}
\caption{The transversal momentum  distribution of final state
hadrons for the two-jet-like
events which is selected at parton level 
(dashed line) and that for unbiased events. }
\label{a6}
\end{figure}




To make sure the observable $SP_t$ is 
really sensitive only for the two-jet-like events we have to 
check if colour connections affect 
the $SP_t$ distribution for unbiased events. 
The answer is yes, see Fig. \ref{a5} (a). 
But $SP_t$ is the sum of the transversal momenta which 
depend on the parameter $\sigma$ in JETSET \cite{lund}, 
which measures the average transversal momentum of 
the quark pair in the string breakings. 
Though $\sigma$ has been 
fitted by experimental groups \cite{newrc}, 
the dependence of $SP_t$ on $\sigma$ 
has never been measured. 
Here we change $\sigma$ to see the dependence of 
$SP_t$ on $\sigma$. Fig. \ref{a5} (b) shows that when 
$\sigma$ is changed from the default value 0.36 
to 0.29, the difference in the $SP_t$ distribution 
for the unbiased events between the 
CS and CC events are nearly smeared. 
In contrast, the global properties of 
the unbiased events are not sensitive 
to this parameter.

It is then critical to check for the two-jet-like events if  the
observable $SP_t$ is sensitive to the parameter $\sigma$.
Comparing Fig. \ref{a4} (b) with (a), we see that the shape of the 
distribution is more or less affected by this parameter, 
but for whatever values of $\sigma$ the difference 
of $SP_t$ between the CS and CC events are always significant.


\begin{figure}
\includegraphics[scale=0.36]{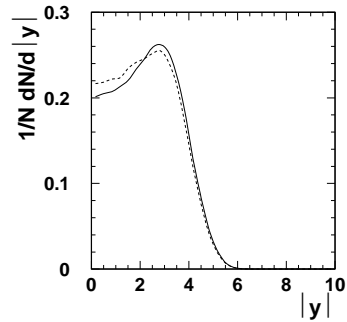}
\caption{The rapidity distribution for the two-jet-like
events selected at the hadronic level.}
\label{a12}
\end{figure}

\begin{figure}
\includegraphics[scale=0.4]{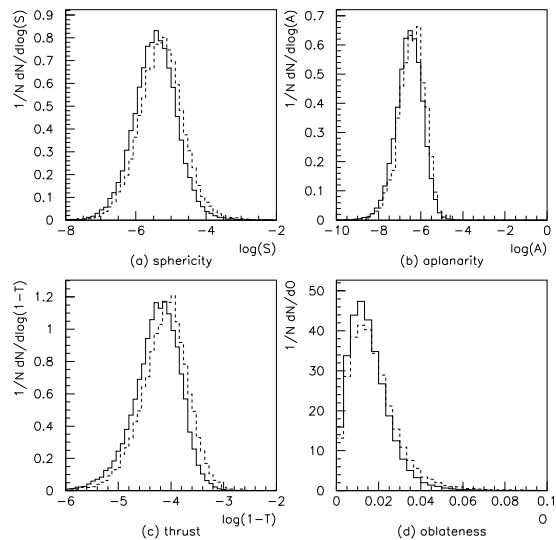}
\caption{Global properties of the two-jet-like
events selected at the hadronic level.}
\label{a13}
\end{figure}

\begin{figure}
\includegraphics[scale=0.36]{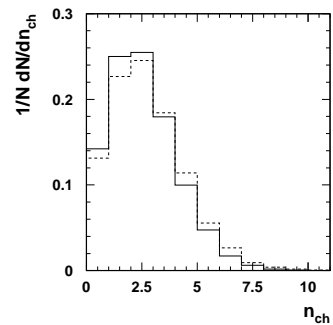}
\caption{$n_{ch}$ distribution with $|y|<0.8$ 
for two-jet-like events selected at the hadronic level. }
\label{a14}
\end{figure}

\begin{figure}
\includegraphics[scale=0.36]{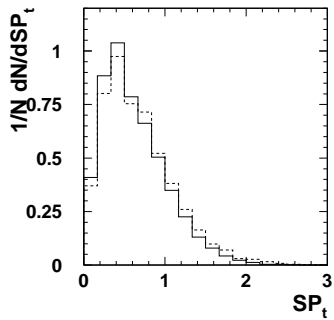}
\caption{$SP_t$ distribution in the rapidity window 
$|y|<0.8$ for the two-jet-like events selected at
the hadronic level. }
\label{a15}
\end{figure}

The proposal in the above 
is at the partonic level which is an ideal situation. 
The events selected by requiring that  
the energy of each gluon must be smaller than a certain value 
cannot be realized in experiments since partons cannot be seen directly.
Now we extend it to the real case, i.e. 
selecting the two-jet-like events according to the criteria 
defined at the hadronic level. 
We notice from Fig. \ref{a6} that the $p_t$ distribution of 
final state hardons (in our case pions, protons and kaons)
for these events is significantly  
different from that for the unbiased events 
in the small $p_t$ region.  
Hence in this work we require that 
the transversal momentum of the hadron 
is smaller than 0.5GeV/c.

These events have similar properties 
to those selected at the partonic level.
This can be seen in Figs. \ref{a12}, \ref{a13}, \ref{a14}.
These results show that the CS two-jet-like events have more particles 
in the small-rapidity region compared to the CC ones. 
The average thrust for the CS two-jet-like 
events is smaller than the CC ones. 
The comparison with Ref. \cite{wangq} shows that the differences in 
the global properties between the CS and CC two-jet-like events are
more obvious than the unbiased events, as is exhibited    
in Fig. \ref{a13}. The most essential fact is that 
$SP_t$ is still a sensitive observable for the way of 
colour connections. One can see this in Fig. \ref{a15} 
which shows there are significant differences  
between the CC and CS two-jet-like events.

Note that our definitions for  
the two-jet-like events and the observable $SP_t$ 
look similar to those in Ref. \cite{eden} at first sight. 
Actually they are different from many aspects. 
Ref. \cite{eden} needed the events with small 
number of gluons to address the issue of 
quark jet hadronization, while in this work 
we prefer those events 
with large number of gluons to make the probability 
of forming the CS states as large as possible. 
Ref. \cite{eden} took the sum of  
$\vec{p}_t$ to demonstrate correlation properties for different
models, while we use the sum of  
the $magitude$ of $\vec{p}_t$ to enlarge the differences between the 
events of different colour connections.


In this paper we extend our previous work on 
the CS states by proposing a possible method to 
probe these types of states in the hadronic process 
of electron-positron annihilation via $Z^0$. 
We find that there would be substantial differences 
in the distribution of $SP_t$ 
in the two-jet-like events for the CS-allowed and for 
the normal colour connections. 
Other distributions, such as those of the 
central rapidity, the $p_t$ and 
the charged particle multiplicity inside 
a central rapidity window, etc., also 
show detectable differences for the CS and the 
CC two-jet-like events. In average there is one 
two-jet-like event in every one thousand hadronic 
events at the $Z^0$ pole. Since large number of events 
have been accumulated at LEP I, 
one can expect to get a good statistics 
in selecting the two-jet-like events to probe 
the presence of the CS states. 
We should mention that the two-jet-like
events defined in this paper and 
the observable $SP_t$ have never been
used before. The future comparison with data for 
this type of events and the $SP_t$ distribution would 
also provide an alternative way of testing different 
hadronization models.

We thank G. Gustafson, Z.-T. Liang and X.-N. Wang for 
insightful discussions. This work is supported in part by the 
National Natural Science Foundation of China (NSFC).


\begin{thebibliography}{99}

\bibitem{lund} T. Sj\"{o}strand, 
Comp. Phys. Commun. {\bf 82}, 74 (1994); 
B. Andersson, G. Gustafson, G. Ingelman and T. Sj\"{o}strand, 
Phys. Rep. {\bf 97}, 31 (1983).

\bibitem{webber} G. Corcella et al., JHEP {\bf 0101}, 010 (2001);
B. R. Webber, Nucl. Phys. B{\bf 238}, 492 (1984); 
G. Marchesini and B. R. Webber, {\it ibid} B{\bf 238}, 1 (1984). 

\bibitem{friberg}  C. Friberg, G. Gustafson, and J. H\"akkinen, 
Nucl. Phys. B{\bf 490}, 289 (1997).

\bibitem{wangqunprd} Q. Wang, G. Gustafson, Y. Jin,  and Q.-B. Xie, 
Phys. Rev. D{\bf 64}, 012006 (2001).

\bibitem{wangq} Q. Wang, G. Gustafson, and Q.-B. Xie, 
Phys. Rev. D{\bf 62}, 054004 (2000).

\bibitem{wangqunold} Q. Wang, Q.-B. Xie, and Z.-G. Si, 
Phys. Lett. B{\bf 388}, 346 (1996); 
Q. Wang and Q.-B. Xie,  Phys. Rev. D{\bf 52}, 1469 (1995).

\bibitem{jinyi} Y. Jin et al., in preparation.

\bibitem{tsj} T. Sj\"ostrand and V. A. Khoze, 
Phys. Rev. Lett. {\bf 72}, 28 (1994); 
Z. Phys. C{\bf 62}, 281 (1994). 

\bibitem{lonnblad} L. L\"onnblad, Z. Phys. C{\bf 70}, 107 (1996).



\bibitem{shaofenglan} F.-L. Shao and Q.-B. Xie, 
High energy Physics and Nuclear Physics (in Chinese) 
{\bf 25}, 710 (2001);  
F.-L. Shao and Q.-B. Xie, to be published in 
High energy Physics and Nuclear Physics (in Chinese). 

\bibitem{bo} B. Andersson, P. Dahlqvist, and G. Gustafson, 
Z. Phys. C{\bf 44}, 461 (1989); {\it ibid} C{\bf 44}, 455 (1989).   

\bibitem{si} Z.-G. Si, Q.-B. Xie, and Q. Wang, 
Commun. Theor. Phys. {\bf 28}, 85 (1997).   

\bibitem{newrc} G. Altarelli, T. Sj\"{o}strand, and Zwirner (Eds.), 
Physics at LEP2, vol.2,  CERN yellow book 96-01. 


\bibitem{eden} P. Ed\'{e}n and G. Gustafson, 
Euro. Phys. Jour. C{\bf 8}, 435 (1999).

\end{thebibliography}
\end{document}